\documentclass[onecolumn,amsmath,amssymb,prd,reprint,longbibliography,floatfix,8pt]{revtex4-1}

\usepackage{cancel}
\usepackage{enumitem}
\usepackage[utf8x]{inputenc}
\usepackage{graphicx}
\usepackage{dcolumn}
\usepackage{bm}
\usepackage[switch]{lineno}
\usepackage{float} 
\usepackage{subfigure}
\usepackage{tikz}
\usepackage{multirow}
\usetikzlibrary{arrows.meta}
\usetikzlibrary{arrows,decorations.pathmorphing,backgrounds,positioning,fit,petri}

\begin{document}

\title{Extending spin-lattice relaxation theory to three-phonon processes}
\author{Nilanjana Chanda, Alessandro Lunghi}
\email{lunghia@tcd.ie}
\affiliation{School of Physics, AMBER and CRANN Institute, Trinity College, Dublin 2, Ireland}

\begin{abstract}
{\bf Spin-lattice relaxation theory has been developed over almost a century, but some cardinal assumptions on the nature of the interactions involved have never been fully verified. This includes the weak coupling approximation, which makes it possible to describe spin dynamics perturbatively and leads to the canonical description of spin relaxation in terms of one- and two-phonon processes. Here, we extend the first-principles theory of spin relaxation to three-phonon processes and apply it to the vdW crystal of a spin-1/2 Chromium nitride complex. Results show that three-phonon contributions to spin relaxation only become relevant at temperatures inaccessible to experiments for this molecule, thus providing unprecedented evidence for the validity of the weak spin-phonon coupling assumption in spin relaxation theory. At the same time, we numerically show that a relatively small increase in spin-phonon coupling would lead to a crossover between three- and two-phonon processes' efficiency at room temperature, illustrating the possibility for three-phonon effects in molecular materials as well as paving the way to a systematic exploration of strong coupling in spin systems.}
\end{abstract}

\maketitle

\section*{Introduction}

In solids, the process of the energy exchange between an electron spin and the lattice vibrations, or phonons, has been of significant importance for nearly a century, particularly in the study of restoring the thermal equilibrium in magnetic resonance experiments, referred to as the spin-lattice or spin-phonon relaxation.
Recently, with the advancement of quantum information science and quantum technologies, a quantitative understanding of spin-phonon relaxation has also become essential across a broad range of magnetic materials and quantum architectures, including molecular nanomagnets\cite{affronte2009molecular}, transition metals \cite{moseley2018spin}, rare-earth ions \cite{orbach1961spin} and solid-state qubits, {\emph e.g.} nitrogen-vacancy (NV) centers in diamond \cite{gugler2018ab, mondal2023spin}, quantum dots \cite{khaetskii2000spin}, and defects in 2D materials \cite{gong2019two, mondal2023spin}.

The first theoretical treatment of spin-lattice relaxation in solids was given by Waller in 1932 \cite{waller1932magnet}. 
He performed the first calculation of relaxation times modulating the dipole-dipole interaction by phonons and discussed the processes of transfer of energy by the spin to the lattice by the one-phonon direct processes as well as the two-phonon Raman processes.
However, Waller found very long relaxation times at low temperature, which did not agree with the experimental results obtained by Gorter in 1936 \cite{gorter1957chapter}.
Around the same time, Heitler \& Teller (1936) \cite{heitler1936time} and Fierz (1938) \cite{fierz1938theory} considered the modulation of the crystalline electric fields by phonons and showed that modulation of the crystalline fields led to a spin-lattice interaction much stronger than that produced by modulations of the magnetic dipolar interaction.
Kronig (1939) \cite{kronig1939mechanism} and Van Vleck (1939, 1940) \cite{van1939jahn, van1940paramagnetic} provided quantitative calculations of the effects of modulation of the crystalline electric field in the iron group alums and obtained quantitative agreement with the experimental results of Gorter (1936) and Teunissen \& Dijkstra (1938) in the high-temperature range, but failed to explain many of their results in the liquid helium range. Building on these early results, Orbach later proposed to employ the theory of spin-lattice relaxation in a phenomonological way \cite{orbach1961spin} rather than attempting to make predictions from first principles and eventually led to the canonical picture of spin relaxation as mediated by either the exchange of one resonant phonon (Direct relaxation), multiple one-phonon processes (Orbach relaxation), and simultaneous two-phonon processes (Raman relaxation).

In parallel to these early attempts to predict spin-lattice relaxation in solid-state magnetic systems, a general theory of open quantum systems was emerging. In 1946, Bloch proposed a set of phenomenological equations in terms of the magnetization, widely known as the Bloch equations, to explain the relaxation process in the context of nuclear magnetic resonance (NMR) \cite{bloch1946nuclear}. Within this formalism, the establishment of thermal equilibrium is determined by two time constants, namely spin-lattice relaxation time $T_1$ and spin-spin relaxation time $T_2$. Motivated by this, Redfield (1957) came up with a microscopic description of the relaxation process and derived an equation of motion for the density operator of the spin system, referred to as a quantum master equation (QME) \cite{redfield1957theory}. The process of relaxation is mathematically represented in the form of a superoperator calculated employing a second-order perturbative treatment on the system-environment interaction of a generic form. 
As such, Redfield's theory can be considered the first formal theory in the study of open quantum system dynamics, including coherent processes. Few years later, in 1976, Gorini, Kossakowski, Sudarshan \cite{gorini1976completely} and at the same time, Lindblad \cite{lindblad1976generators} independently came up with the most general form for the generator of a quantum dynamical semigroup, known as the GKSL form which defines a Markovian QME.
This has been used as a standard result for both microscopic and phenomenological studies in open quantum systems \cite{alicki2007quantum,kraus2008preparation,sieberer2016keldysh,wang2011quantum}.

However, the Born-Markov or weak-coupling approximations may break down in many realistic cases, \emph{e.g.} when the timescale separation between the system and environment breaks down, for strong system-environment coupling, large initial system-environment correlations, structured or finite reservoirs, low temperatures, etc. \cite{breuer2016colloquium}. In such cases, more general tools are required. The formally exact description of open quantum systems dynamics is usually studied through two distinct approaches: the time-nonlocal Nakajima-Zwanzig quantum master equations (NZ QMEs) \cite{nakajima1958quantum, zwanzig1960ensemble}; and the time-convolutionless, or time-local, quantum master equations (TCL QMEs) \cite{shibata1977generalized, chaturvedi1979time, shibata1980expansion}, as expressed by
\begin{eqnarray}
    \frac{d\hat{\rho}(t)}{dt}=\hat{\hat{R}}(t,t_0)\hat{\rho}(t)\:.
    \label{TCL}
\end{eqnarray}

Even without leaving the Markovian regime, QMEs in the TCL formalism can be used to study open quantum systems dynamics through a perturbative expansion of the generator $\hat{\hat{R}}(t,t_0)$ with respect to the system-environment coupling. Using this method, the relaxation and decoherence processes in any quantum system, not necessarily weakly coupled to its environment, can be explained. A wide range of validity of this method has been discussed in detail in \cite{breuer1999stochastic} by means of a damped Jaynes-Cummings model. In particular, the fourth order of the TCL QMEs is shown to be a good approximation to the exact solution for a large range of physically relevant parameters \cite{breuer2001time}. This perturbative treatment is often applied to spin systems interacting with a bath of harmonic oscillators, termed spin-boson models \cite{leggett1987dynamics}
and thus forming an ideal theoretical basis for discussing spin-phonon relaxation in molecular crystals or other solid-state systems such as defects or impurities.

In recent years, a similar approach to TCL, has indeed been used to systematically merge open quantum systems theory and first-principles methods with the aim of completing that process initiated almost 100 years ago by Waller, Kronig and Van Vleck \cite{lunghi2019phonons,lunghi2020limit,lunghi2022toward, mariano2025role}. Such formalism, known as $T$-matrix formalism, corresponds to a simplification of full TCL QMEs \cite{schieve2009quantum} in retaining only terms appearing as a generalized Fermi's golden rule \cite{bruus2004many}. Once these QMEs are combined with electronic structure methods, all the coefficients required to implement the former for specific molecules can be specified, thereby ensuring a fully non-parametric treatment of all the relevant interactions, including spin and phonon spectra as well as the spin-phonon coupling coefficients. Both one- and two-phonon processes following second- and fourth-order perturbative treatment are now well-characterized \cite{lunghi2019phonons,lunghi2020limit,lunghi2022toward, mariano2025role}.

In this work, we adopt the $T$-matrix formalism, which systematically incorporates spin-phonon transitions at all possible orders, and expand existing efforts to include the contribution of three-phonon processes in spin-lattice relaxation. To quantify the relevance of these terms, we build on the recent success of ab initio methods to predict spin-phonon relaxation in solid-state spin-1/2 molecular materials and numerically implement the theory for a Cr(V) coordination compound \cite{mariano2025role}. Molecules of this class are under intense scrutiny as molecular qubits, and key questions about their mechanism of relaxation at room temperature are still debated in the literature \cite{eaton2025anisotropy,kazmi2025esr}, thus offering a good opportunity to benchmark theoretical and computational efforts. Results show that, for such a system, two-phonon relaxation theory is well justified, but strong-coupling effects might become effective for a relatively small increase in spin-phonon coupling.

\section*{$T$-matrix theory for spin relaxation}

\subsection*{The general theory} 

The transfer-matrix or $T$-matrix formalism was originally introduced in scattering theory and has been developed to describe tunneling processes. This framework defines an operator $\hat{T}$ that transforms a free scattering state into an interacting scattering state. 
The $T$-matrix theory has been widely used in the context of condensed matter physics \cite{akera1999coulomb,golovach2004transport} and nuclear physics \cite{gross1999covariant,soma2008medium} over the years.
Later on, $T$-matrix formalism has been extended to explain the tunneling and relaxation processes taking place in molecular systems \cite{koch2004thermopower,jorn2006theory}.
Here, we briefly outline the derivation of the $T$-matrix formula using time-dependent perturbation theory to calculate the transition rates under this framework and draw its connection with the exact TCL generator.

Let us consider a quantum system described by the following total Hamiltonian,
\begin{equation}
   \hat{\mathcal{H}} = \hat{H_0} + \hat{H}(t)
\end{equation}
where $\hat{H}_0$ denotes the Hamiltonian of the unperturbed system and $\hat{H}(t)$ denotes a perturbation adiabatically turned on at the distant past, {\emph i.e.} $t_0 \rightarrow -\infty $, such that
\begin{equation}
    \hat{H}(t) = \hat{H} e^{\eta t} \:,
\end{equation}
with the rate $\eta$ being a small and positive number.
The choice of such an adiabatic switch-on implies that at the initial time $t_0 = -\infty $, the system and the environment are decoupled. 

The time evolution of quantum systems from time $t_0$ to $t$ is defined in terms of a unitary propagator $\hat{U} (t,t_0)$,
\begin{equation}
\label{psi-t}
|\psi (t) \rangle = \hat{U}(t,t_0) \ |\psi (t_0) \rangle.
\end{equation}
Inserting Eq. \ref{psi-t} in the Schr\"odinger equation (expressed in the interaction picture), one can write,
\begin{equation}
\label{Sch-eq-U}
\frac{d}{dt} \hat{U}(t, t_0) = - \frac{i}{\hbar} \ \hat{H}^I (t) \ \hat{U}(t, t_0).
\end{equation}
where $\hat{H}^I (t) =  e^{i \hat{H}_0 t/ \hbar}  \ \hat{H}(t) \ e^{-i \hat{H}_0 t/ \hbar } $.
Eq. \ref{Sch-eq-U} can be solved for $ \hat{U}(t, t_0)$ iteratively in the interval $[t_0, t]$, leading to the conventional Dyson series,
\begin{eqnarray}
\label{U-tn} \hat{U}(t, t_0)
=&& \sum_{n=0}^\infty \frac{1}{n!}
\left( - \frac{i}{\hbar} \right)^n \int_{t_0}^t dt_1 \int_{t_0}^{t}  dt_2 \dots \int_{t_0}^{t}  dt_n \nonumber \\ 
&& \mathcal{T} [ \hat{H}^I(t_1) \hat{H}^I(t_2) \dots \hat{H}^I(t_n)]   \:,
\end{eqnarray}
where $\mathcal{T}$ denotes the time-ordering operator. This expression of $\hat{U}(t, t_0) $ in Eq. \ref{U-tn} is considered as the starting point for infinite-order perturbation theory.
 
The probability to observe a transition from the initial state $|i \rangle$ to the final state $|f \rangle$, both eigenstates of $\hat{H}_0$, at time $t$ can be expressed as (the detailed derivation can be found in the Supplementary Materials),
\begin{eqnarray}
P_{fi} &=& |	\langle f| i(t) \rangle |^2 =\nonumber \\
\label{prob-tn} &=&
\frac{ e^{2 \eta t}}{[(E_f - E_i)^2 + (\eta \hbar)^2 ]}
\left| \langle f| \hat{T} | i \rangle \right|^2 \:,
\end{eqnarray}
where the operator $\hat{T}$ is defined as,
\begin{eqnarray}
\label{T-matrix-def}
\hat{T}
= \sum_{n=1}^\infty \hat{H} \ \left[ \frac{1}{(E_i - \hat{H}_0 + i \hbar \eta)} \hat{H} \right]^{n - 1}
= \sum_{n=1}^\infty \hat{T}^{(n)} .
\end{eqnarray}
Following Eq. \ref{prob-tn}, the transition rate between the states $|i\rangle$ and $|f\rangle$ for $\eta \rightarrow 0^+ $ can be expressed as
\begin{eqnarray}
\label{gen-Frule}
R_{fi} = \frac{d P_{fi}}{dt}  = \frac{2 \pi}{\hbar} \left| \langle f |\hat{T} | i \rangle 
\right|^2 \delta(E_f - E_i)\:,
\end{eqnarray}
where $E_i$ and $E_f$ are the eigenenergies of the states $|i\rangle$ and $|f\rangle$, respectively. 
The expression shown in Eq. \ref{gen-Frule} defines a time-independent (Markovian) generator for the diagonal elements of the density matrix as
\begin{equation}
    \dot{\rho}_{ff}= R_{fi} \: \rho_{ii}(0) \sim R_{fi} \: \rho_{ii}(t)\:,
    \label{Tmat}
\end{equation}
where the approximation made in the r.h.s of Eq. \ref{Tmat} is exactly the difference between the $T$-matrix approximation to the dynamics of open quantum systems and the exact TCL formalism \cite{timm2011time}. The discussion of the validity of such an approximation is beyond the scope of the present work and will be addressed separately.

The expression given in Eq. \ref{gen-Frule} reflects a generalized Fermi's golden rule, which can be used to define the generator of the QME at any perturbative order, namely
\begin{equation}
\label{R-series}
    R_{fi} = \sum_{n=1}^\infty R^{(2n)}_{fi} \:.
\end{equation}
We note that the appearance of the generator exclusively in even orders in Eq. \ref{R-series} is a consequence of assuming that the perturbation has zero expectation value, as is the case for linear spin-phonon interactions. While expressions for the first two orders, $n=1, 2$, have already been unravelled \cite{lunghi2019phonons,lunghi2022toward}, here we tackle the case for $n=3$.

\subsection*{Spin-phonon coupling}

Let us now explicitly consider the interaction between an electron spin and its surrounding phonon bath originating from the thermal dynamics of the vdW molecular crystal. In the following, we will refer to the system as a spin, but the formalism effectively applies to any quantum system interacting with a bosonic environment. The total Hamiltonian is expressed as
\begin{equation}
\hat{\mathcal{H}} = \hat{H}_0 + e^{\eta t} \hat{H}_{\rm sph}\:,
\end{equation}
where $\hat{H}_0 = \hat{H}_s + \hat{H}_{\rm ph} $ stands for the decoupled spin-phonon system, with the spin Hamiltonian $\hat{H}_s $ describing the low-lying electronic states of the molecular
systems and the phonon Hamiltonian $\hat{H}_{\rm ph} $ being considered to be a collection of quantum harmonic oscillators. The coupling between the spin and the phonon bath, $\hat{H}_{\rm sph}$ represents the perturbation in the system, which, under weak-coupling approximation, is considered as a Taylor expansion of $\hat{H}_s $ with respect to the atomic displacements associated with the phonons $Q_\alpha=( a_\alpha^\dagger + a_\alpha )$ up to the first order as follows, 
\begin{equation}
\label{Hsph-def}
\hat{H}_{\rm sph} 
= \sum_\alpha \left( \frac{\partial \hat{H}_s}{\partial  Q_\alpha} \right)  Q_\alpha 
= \sum_\alpha V^\alpha Q_\alpha \:.
\end{equation}
The eigenstates of $\hat{H}_s$ are assumed to be non-degenerate and are referred to as $|a\rangle$, and the eigenstates of $\hat{H}_{\rm ph} $ are referred to as $|v \rangle$, which will correspond to any of the possible occupations $|n_\alpha \, n_\beta \dots \rangle$ of each vibrational degree of freedom. 
Therefore, the eigenstates of $\hat{H}_0$ of the composite system (spin + phonon), will be expressed as $|i\rangle = |a \ n_\alpha \, n_\beta \dots \rangle $. Results can be generalized to the full density matrix and degenerate \cite{lunghi2025fourth}, but here we will focus on the study of the time evolution of the diagonal elements of the density matrix spanning the states of $\hat{H}_0$ following the $T$-matrix formalism detailed in the previous section by assuming that $\hat{H}_{\rm sph} $ acts as a perturbation. 

\subsection*{Second-order generator}

Let us now set $n = 1$ in the expression of $T$-matrix in Eq. \ref{T-matrix-def} and derive an expression for the second-order contribution to the generator given in Eq. \ref{R-series}. For $n=1$, the $T$-matrix simply reads 
\begin{equation}
\label{T-matrix-2nd}
\hat{T}^{(1)} =  \hat{H}_{\rm sph} \:, 
\end{equation}
leading to a second-order generator given by
\begin{equation}
\label{G-2nd}
 R^{(2)}_{fi} = \frac{2 \pi}{\hbar} \left| \langle f | \hat{H}_{\rm sph} | i \rangle \right|^2 \delta(E_f - E_i) .
\end{equation}

Since the annihilation (creation) operators, $a_\alpha$
($a_\alpha^\dagger$), only act once between states $|i\rangle$ and $|f\rangle$, transitions will only occur among states that differ by one $\alpha$-phonon occupation number. We will therefore consider an initial state 
\begin{equation}
\label{ini-2nd}
|i \rangle = | a \ n_\alpha \rangle.
\end{equation}
and final states
\begin{align}
\label{fstates-2nd}
|f^{+} \rangle = & | b \ (n_\alpha + 1) \rangle \\ 
|f^{-} \rangle = & | b \ (n_\alpha - 1) \rangle \:,
\end{align}
with the exclusion of $i=f$, which is treated separately.

\begin{figure}[h!]
\includegraphics[width=\linewidth]{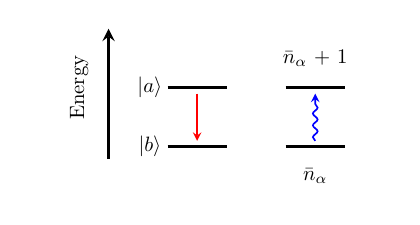}
    \caption{A spin transition from the state $| a\rangle$ to $| b \rangle$ (red solid line), due to a resonant one-phonon emission (blue wavy line) process as dictated by the generator $R^{(2)}_{ba} $.}
    \label{1-ph-fig}
\end{figure}

Let us first consider the case $|f^{+} \rangle = | b \ (n_\alpha + 1) \rangle$. 
The only non-vanishing contribution of $\hat{H}_{\rm sph}$ associated to this final state is $ V^\alpha  a_\alpha^\dagger$. Inserting Eqs. \ref{fstates-2nd} and \ref{ini-2nd} in Eq. \ref{G-2nd}, and taking a statistical average over all possible initial phonon states $|n_\alpha\rangle$, we obtain the transition rate for the final state $|f^{+} \rangle$ as
\begin{eqnarray}
\label{W-1ph}
R^+_{ba} =\frac{2\pi}{\hbar^2}  \sum_\alpha  \left| V^\alpha_{ba} \right|^2 
G_{+} (\omega_{ba}, \omega_\alpha) \:,
\end{eqnarray}
where $ V^\alpha_{ba}$ denotes the matrix element of the linear coupling operator for the phonon mode $\alpha$ among the eigenstates $ | b \rangle$ and $ | a \rangle$ of $\hat{H}_s$, and $G_{+} (\omega_{ba}, \omega_\alpha) = (\bar{n}_\alpha + 1) \ \delta (\omega_{ba} + \omega_\alpha) $
with
$\hbar\omega_{ba} = (E_b - E_a) $ being the energy spacing between the spin states
and $\bar{n}_\alpha = [\exp(\hbar \omega_\alpha/{\rm k_B} T) - 1]^{-1}$ denoting the thermal phonon population following Bose-Einstein distribution. The energy conservation in the process is confirmed by the presence of the Dirac delta function in the expression of the transition rate. The physical picture that corresponds to the transition rate given in Eq. \ref{W-1ph} is schematically represented in Fig. \ref{1-ph-fig}, which 
reports a spin making a transition from the higher energy state $| a \rangle$ to the lower energy state $| b \rangle$ while one phonon is emitted into the lattice. 

Similarly, for the case $|f^{-} \rangle = | b \ (n_\alpha - 1) \rangle$, we note that the only non-vanishing contribution of $\hat{H}_{\rm sph}$ to the associated transition is $ V^\alpha  a$.
Inserting Eqs. \ref{fstates-2nd} and \ref{ini-2nd} in Eq. \ref{G-2nd}, and performing the same average as before, we obtain the transition rate 
\begin{equation}
R^-_{ba} = \frac{2\pi}{\hbar^2} \sum_\alpha \left| V^\alpha_{ba} \right|^2  
G_{-} (\omega_{ab}, \omega_\alpha) \:,
\end{equation}
where $G_{-} (\omega_{ab}, \omega_\alpha) = \bar{n}_\alpha \ \delta (\omega_{ab} - \omega_\alpha) $.

The total transition rate of a spin transition from the state $| a\rangle$ to $| b \rangle$ due to the second-order generator can be determined by taking the sum of all one-phonon processes, i.e.
\begin{equation}
R^{(2)}_{ba} = R^+_{ba} + R^-_{ba} .
\end{equation}

\subsection*{Fourth-order generator}

Let us now set $n = 2$ in the expression of $T$-matrix in Eq. \ref{T-matrix-def} and derive an expression for the fourth-order contribution to the generator given in Eq. \ref{R-series}. 
For $n=2$, the $T$-matrix can be calculated as  
\begin{eqnarray}
\label{T-matrix-4th}
\hat{T}^{(2)} 
= \sum_{l} \hat{H}_{\rm sph}  \left[ \frac{|l\rangle \langle l|}{(E_i - E_l + i 0^+)}  \hat{H}_{\rm sph}
\right] \:,
\end{eqnarray}
where a complete basis $\{ |l\rangle \}$ is introduced in between the operators (as shown in the Supplementary Materials), leading to a fourth-order generator given by
\begin{eqnarray}
\label{G-4th}
R^{(4)}_{fi} 
= \frac{2 \pi}{\hbar} \left| 
\sum_{l} \frac{\langle f | \hat{H}_{\rm sph} |l \rangle \langle l| \hat{H}_{\rm sph} | i \rangle}{(E_i - E_l + i 0^+)}  \right|^2 
\delta(E_f - E_i).
\end{eqnarray}
Since $\hat{H}_{\rm sph}$ now acts twice between the initial and final state, we will consider transitions that involve a change of occupation numbers for two phonons, $\alpha$ and $\beta$. Therefore, for a general initial state
\begin{equation}
\label{ini-4th}
|i \rangle = | a \ n_\alpha n_\beta \rangle \:,
\end{equation}
there will be four possible final states: $|f^{++} \rangle $, $|f^{--} \rangle$, $|f^{+-} \rangle$, $|f^{-+} \rangle$ (details in the Supplementary Materials).

\begin{figure}[h!]
\includegraphics[width=\linewidth]{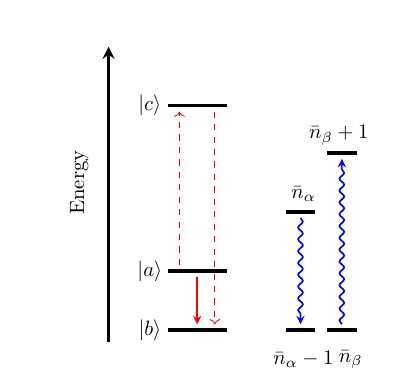}
    \caption{A spin transition from the state $| a\rangle$ to $| b \rangle$ (red solid line), due to a two-phonon process as dictated by the superoperator $ R^{(4)}_{ba} $, where one phonon is absorbed and one is emitted simultaneously (blue wavy lines), and the entire process is mediated through intermediate virtual transitions (red dashed lines) to an excited spin state $| c \rangle$.
    }
    \label{2-ph-fig}
\end{figure}

Let us consider the case $|f^{-+} \rangle = | b \ (n_\alpha - 1) (n_\beta + 1) \rangle$. 
For $\alpha \neq \beta $, there are only two possible intermediate states which possess a non-zero contribution to the sum over $l$, 
\begin{align}
|l \rangle = & | c \ (n_\alpha - 1) n_\beta  \rangle \:, \\
|l \rangle = & | c \ n_\alpha (n_\beta + 1)  \rangle \:.
\end{align}
Inserting Eq. \ref{ini-4th} in Eq. \ref{G-4th}, 
and taking a statistical average over all possible initial phonon states $|n_\alpha n_\beta \rangle$, we obtain the transition rate for the final state $|f^{-+} \rangle$ as
\begin{equation}
\label{W-2ph}
R^{-+}_{ba}  
= \frac{2\pi}{ \hbar^2} \sum_{\alpha < \beta} \left| T_{ba}^{\beta \alpha, -} + T_{ba}^{\alpha \beta, +}
\right|^2 \ G_{-+} (\omega_{ba}, \omega_\alpha, \omega_\beta)  .
\end{equation}
where 
\begin{equation}
T_{ba}^{\alpha \beta, \pm} 
= \sum_{c}  \frac{ V^\alpha_{bc}  V^\beta_{ca} }{(E_c - E_a \pm \hbar \omega_\beta)}
\end{equation}
with $G_{-+} (\omega_{ba}, \omega_\alpha, \omega_\beta) = \bar{n}_\alpha (\bar{n}_\beta +1)  \ \delta(\omega_{ba} - \omega_\alpha + \omega_\beta )$.

Fig. \ref{2-ph-fig} schematically represents the process underlying Eq. \ref{W-2ph}, where a spin transition from the state $| a \rangle$ to $| b \rangle$ is mediated by virtual transitions to/from an intermediate state $| c \rangle$ and a simultaneous absorption and emission of two different phonons. It is important to remark that this is an irreducible two-phonon Raman process, meaning that although the intermediate state $|c\rangle$ is a real eigenstate of $\hat{H}_{\rm s}$, it never gets populated.

Finally, the total transition rate of a spin transition from the state $| a\rangle$ to $| b \rangle$ due to the fourth-order generator can be determined by taking the sum of all two-phonon processes, i.e.
\begin{eqnarray}
R^{(4)}_{ba} =  R^{++}_{ba}  + R^{--}_{ba} + R^{+-}_{ba}   + R^{-+}_{ba} \:,  
\end{eqnarray}
representing double phonon absorption, double phonon emission, and absorption/emission processes, respectively. The full expression of $R^{(4)}_{ba}$ is reported in the Supplementary Materials.

\subsection*{Sixth-order generator}

Let us now set $n = 3$ in the expression of $T$-matrix in Eq. \ref{T-matrix-def} and derive an expression for the sixth-order contribution to the generator given in Eq. \ref{R-series}. 
For $n=3$, the $T$-matrix can be calculated as 
\begin{align}
\label{T-matrix-6th}
 & \hat{T}^{(3)} = \\
 & \sum_{ln} \hat{H}_{\rm sph} 
\left[ \frac{|l\rangle \langle l|}{(E_i - E_l + i 0^+)}   \hat{H}_{\rm sph} \right] \left[ \frac{|n\rangle \langle n|}{(E_i - E_n + i 0^+)}  \hat{H}_{\rm sph} \right] \nonumber
\end{align}
leading to a sixth-order generator given by
\begin{align}
\label{G-6th}
& R^{(6)}_{fi} 
= \\
& \frac{2 \pi}{\hbar} \left| 
\sum_{ln} \frac{\langle f | \hat{H}_{\rm sph} |l \rangle \langle l | \hat{H}_{\rm sph} |n \rangle 
\langle n| \hat{H}_{\rm sph} | i \rangle}{(E_i - E_l + i 0^+) (E_i - E_n + i 0^+)}  \right|^2  \delta(E_f - E_i).  \nonumber
\end{align}

Following the same logic as in previous sections, we now pick a general initial state with three different phonons,
\begin{equation}
\label{ini-6th}
|i \rangle = | a \ n_\alpha n_\beta n_\gamma \rangle \:,
\end{equation}
which will lead to transitions to states that differ by one $\alpha$-phonon, one $\beta$-phonon, and one $\gamma$-phonon occupation numbers. Following the notation introduced previously, there are eight such final states possible: $|f^{+++} \rangle $, $|f^{---} \rangle$, $|f^{+--} \rangle$, $|f^{-+-} \rangle$, $|f^{--+} \rangle$, $|f^{-++} \rangle$, $|f^{+-+} \rangle$, $|f^{++-} \rangle$, as detailed in the Supplementary Materials.

Let us consider the case $|f^{-++} \rangle = | b \ (n_\alpha - 1) (n_\beta + 1) (n_\gamma + 1) \rangle$. 
For $\alpha \neq \beta \neq \gamma $, there are  six possible combinations of the states $|l \rangle$ and $|n \rangle$ which form the secular pairs having non-zero contribution to the sum over $l, n$, i.e.
\begin{align}
|l \rangle = & | c \ (n_\alpha - 1) (n_\beta +1) n_\gamma  \rangle,
\ \ |n \rangle = | d \ n_\alpha (n_\beta +1) n_\gamma  \rangle \nonumber \\
|l \rangle = & | c \ (n_\alpha - 1) (n_\beta +1) n_\gamma  \rangle,
\ \ |n \rangle = | d \ (n_\alpha - 1) n_\beta n_\gamma  \rangle \nonumber \\
|l \rangle = & | c \ (n_\alpha - 1) n_\beta  (n_\gamma +1) \rangle,
\ \ |n \rangle = | d \ (n_\alpha - 1) n_\beta n_\gamma  \rangle \nonumber \\
|l \rangle = & | c \ (n_\alpha - 1) n_\beta (n_\gamma +1) \rangle,
\ \ |n \rangle = | d \ n_\alpha n_\beta  (n_\gamma +1)  \rangle \nonumber \\
|l \rangle = & | c \ n_\alpha (n_\beta +1) (n_\gamma +1) \rangle,
\ \ |n \rangle = | d \ n_\alpha (n_\beta +1) n_\gamma  \rangle \nonumber \\
|l \rangle = & | c \ n_\alpha (n_\beta +1) (n_\gamma +1) \rangle,
\ \ |n \rangle = | d \ n_\alpha n_\beta  (n_\gamma +1) \rangle . \nonumber
\end{align}
Inserting Eq. \ref{ini-6th} in Eq. \ref{G-6th}, 
and taking a statistical average over all possible initial phonon states $|n_\alpha n_\beta n_\gamma \rangle$, we obtain the transition rate for the final state $|f^{-++} \rangle$ as
\begin{widetext}
\begin{equation}
\label{W-3ph}
\kern-8mm  R^{-++}_{ba}  
=  \frac{2 \pi}{\hbar^2} \sum_{\alpha < \beta < \gamma} \left| 
T_{ba}^{\gamma \beta \alpha, +-}
+ T_{ba}^{\gamma \alpha \beta, -+}
+ T_{ba}^{\beta \gamma \alpha, +-}
+T_{ba}^{\beta \alpha \gamma, -+}
+ T_{ba}^{\alpha \gamma \beta, ++}
+ T_{ba}^{\alpha \beta \gamma, ++}
\right|^2
G_{-++} (\omega_{ba}, \omega_\alpha, \omega_\beta, \omega_\gamma)
\end{equation}
\end{widetext}
where 
\begin{equation}
    T_{ba}^{\alpha \beta \gamma, \pm \pm} 
= \sum_{cd}  \frac{ V^\alpha_{bc}  V^\beta_{cd} V^\gamma_{da} }{(E_c - E_a \pm \hbar \omega_\beta \pm \hbar \omega_\gamma )  (E_d - E_a \pm \hbar \omega_\gamma )} \:,
\end{equation}
and 
\begin{align}
    & G_{-++} (\omega_{ba}, \omega_\alpha, \omega_\beta, \omega_\gamma) = \nonumber \\ 
    & \bar{n}_\alpha (\bar{n}_\beta +1)  (\bar{n}_\gamma +1) \ \delta(\omega_{ba} - \omega_\alpha + \omega_\beta + \omega_\gamma)  \:.
\end{align}

\begin{figure}[h!]
   \includegraphics[width=\linewidth]{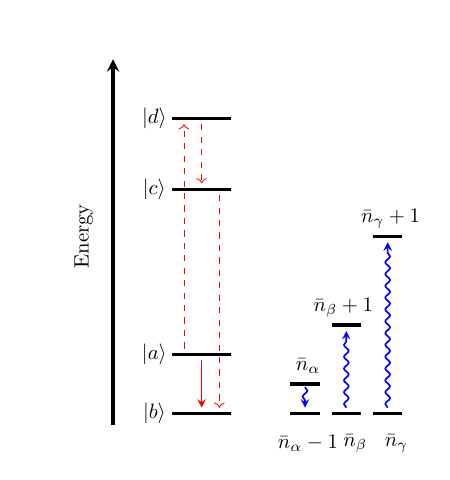}
    \caption{A spin transition from the state $| a\rangle$ to $| b \rangle$ (red solid line), due to a three-phonon process as dictated by the superoperator $ R^{(6)}_{ba}  $, where one phonon is absorbed and two are emitted simultaneously (blue wavy lines), and the entire process is mediated through intermediate virtual transitions (red dashed lines) to the excited spin states $| c \rangle$ and $| d \rangle$.}
    \label{3-ph-fig}
\end{figure}

The expression of $R^{-++}_{ba}$ dictates a transition from the spin state $| a \rangle$ to $| b \rangle$ induced by the simultaneous interaction of the spin with three phonons in the states $| n_\alpha \rangle$, $| n_\beta \rangle$, $| n_\gamma \rangle$, where one is absorbed ($ n_\alpha $) by the spin and two are emitted ($ n_\beta $, $ n_\gamma $). 
The energy difference between the three phonons is the same as the energy spacing between the spin states, which is reflected by the presence of the Dirac delta function.
Fig. \ref{3-ph-fig} schematically reports such a three-phonon process, including the three intermediate virtual transitions involving a pair of eigenstates of $\hat{H}_s$, $|c\rangle$ and $|d\rangle$.

The total sixth-order generator for a transition between the state $| a\rangle$ and $| b \rangle$ is then determined by taking the sum of all three-phonon processes (all possible $| f\rangle$ states as detailed in the Supplementary Materials), i.e.
\begin{align}
\label{R6-total}
      R^{(6)}_{ba} = & R^{+++}_{ba}  + R^{---}_{ba}  + R^{+--}_{ba} + R^{-++}_{ba} + \nonumber \\ 
      & R^{-+-}_{ba} + R^{+-+}_{ba} + R^{--+}_{ba} + R^{++-}_{ba} . 
\end{align}

\section*{Ab initio implementation for a crystal of magnetic molecules}

\begin{figure}[h]
\includegraphics[width=\linewidth]{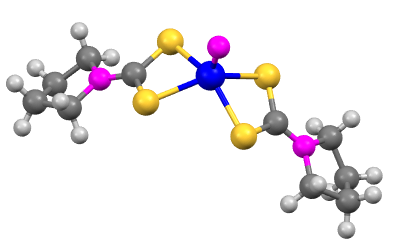}
 \caption{Molecular structure of CrN(pyrdtc)$_{\mathrm 2}$. Color code: blue for Cr, magenta for N, dark gray for C, yellow for S, light gray for H}
    \label{fig-mol}
\end{figure}

The theoretical results obtained in the previous section are here numerically implemented at the first principles level for the crystal of nitrido bis(pyrrolidine dithiocarbamate)chromium(V) [CrN(pyrdtc)$_{\mathrm 2}$] \cite{,birk2003atom}. The molecular structure of this compound, reported in Fig. \ref{fig-mol}, includes a Cr(V) ion in a square pyramidal coordination geometry and behaves as a spin-1/2 due to the presence of a single unpaired electron localized in the d$_{\mathrm{xy}}$ orbital of Cr. Due to the strong ligand field exerted on the d orbitals of Cr, the first d-d electronic excitations are about 30,000 cm$^{-1}$, which confer a rather isotropic nature to the spin-1/2 ground state of this molecule \cite{mariano2025role}. Following previous work \cite{mariano2025role}, here we interpret the system Hamiltonian $\hat{H}_s$ as the full electronic molecular Hamiltonian, and not as an effective spin-1/2 Hamiltonian. Under these conditions, the system's Hilbert space spans ten electronic states, each corresponding to a single electron located in one of the five d orbitals of Cr with either spin up or down, and the system-environment coupling is represented by vibronic coupling \cite{mariano2024vibr}. The electronic structure, phonons and vibronic coupling coefficients of CrN(pyrdtc)$_{\mathrm 2}$ were computed ab initio in a previous work \cite{mariano2025role}, and here we employ such a dataset to perform our simulations. 

Before turning to the simulation of spin relaxation, we first must note that one-phonon processes mediated by $R^{(2)}$ are not relevant for such a system. Energy conservation would require phonons to either have an energy comparable to the ground-state doublet Zeeman splitting (fractions of cm$^{-1}$), or commensurate with the energy separation between the electronic ground state and the electronic excited states (above 30,000 cm$^{-1}$). In the latter case, phonons with such energies are not available as vibrational spectra do not extend beyond about 3,000 cm$^{-1}$. Phonons of very low energy, in principle, exist in a crystal and correspond to acoustic modes with very long wavelength \cite{lunghi2019phonons}. However, these phonons will only sensibly contribute to spin relaxation at very low temperatures \cite{lunghi2020limit}. Here, we instead focus on high temperatures, where strong-coupling effects might arise and work at the $\Gamma$-point level of approximation, therefore only retaining optical phonons in our simulations.

\begin{figure}[h!]
    \includegraphics[width=\linewidth]{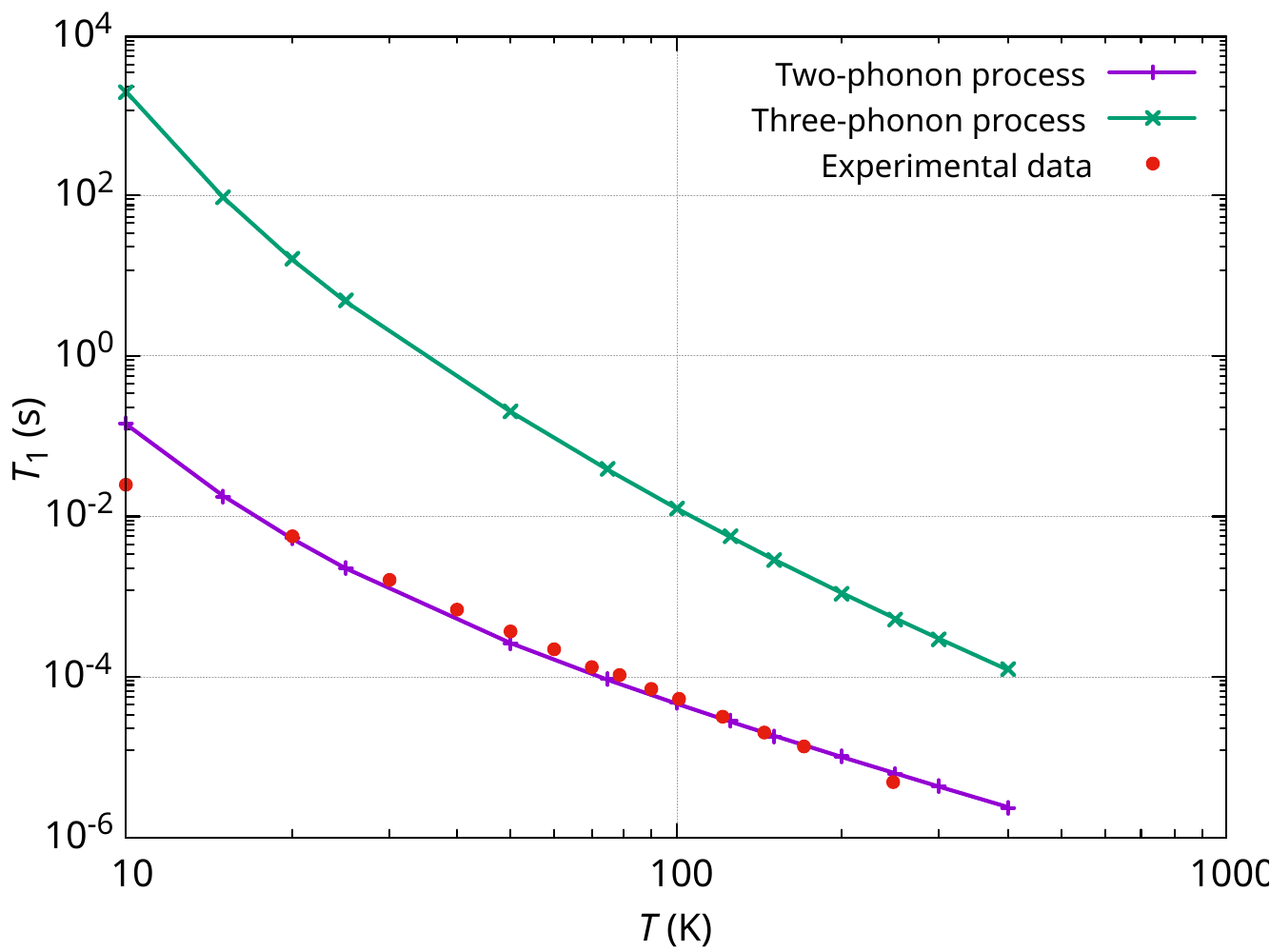}
    \caption{$T_1$ computed with three-phonon (green line) as well as two-phonon (purple line) processes is reported for the temperature range $T =$ 5-400 K. The two-phonon contributions show a good agreement with the experimental data (red points).}
    \label{fig-T1}
\end{figure}

Two-phonon contributions to spin relaxation, as introduced by $R^{(4)}$, for CrN(pyrdtc)$_{\mathrm 2}$ have been studied previously by means of both electron paramagnetic resonance and ab initio simulations, and a generally good agreement between the two was found \cite{mariano2025role}. Here we reproduce the latter results and report them in Fig. \ref{fig-T1}

We can now turn our attention to the implementation of $R^{(6)}$, which accounts for three-phonon processes. The latter are now implemented in the development branch T4 of the software MolForge \cite{lunghi2022toward}. The values of $T_1$ obtained under the effect of such contributions to spin relaxation are reported as a function of temperature in Fig. \ref{fig-T1}. In the high-temperature limit, $T_1$ assumes a power law dependency with respect to the temperature as $T^{-3}$, as expected from the product of three Bose-Einstein factors. Despite this rapid increase in relaxation efficiency with temperature, two-phonon effects remain the most relevant contributions well above room temperature, confirming the accuracy of previous simulations \cite{mariano2025role}.

\begin{figure}[h!]
\includegraphics[width=\linewidth]{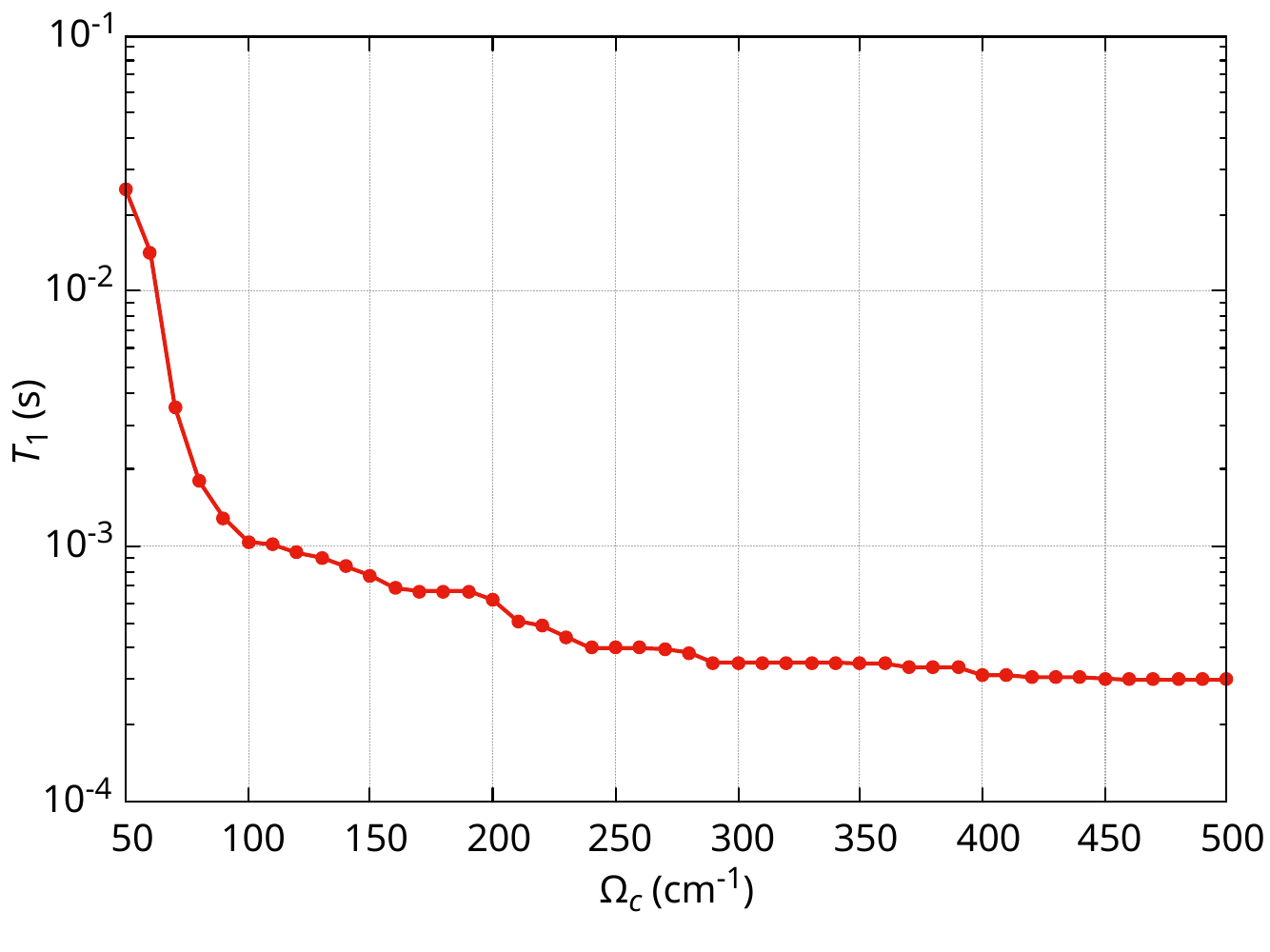}
   \caption{Phonon energy convergence. The value of $T_1$ computed with three-phonon process at 300 K is reported with a red line and dots as a function of the phonons high-energy cutoff $\hbar\Omega_c$.}
    \label{fig-T1-vs-wc}
\end{figure}

We nonetheless provide a more in-depth analysis of three-phonon contributions, and perform the simulations of $T_1$ by including all phonons up to an energy cutoff $\hbar\Omega_c$. Fig. \ref{fig-T1-vs-wc} reports the dependency of $T_1$ computed from $R^{(6)}$ as a function of $\Omega_c$ and shows that while a drastic reduction occurs by including phonons in the range 50-100 cm$^{-1}$, a full convergence is only achieved by introducing phonons at energies up to at least 300 cm$^{-1}$. This is in contrast to two-phonon processes, which were found to be dominated by the phonons in the lower energy window \cite{mariano2025role}. 

\begin{figure}[h!]
\includegraphics[width=\linewidth]{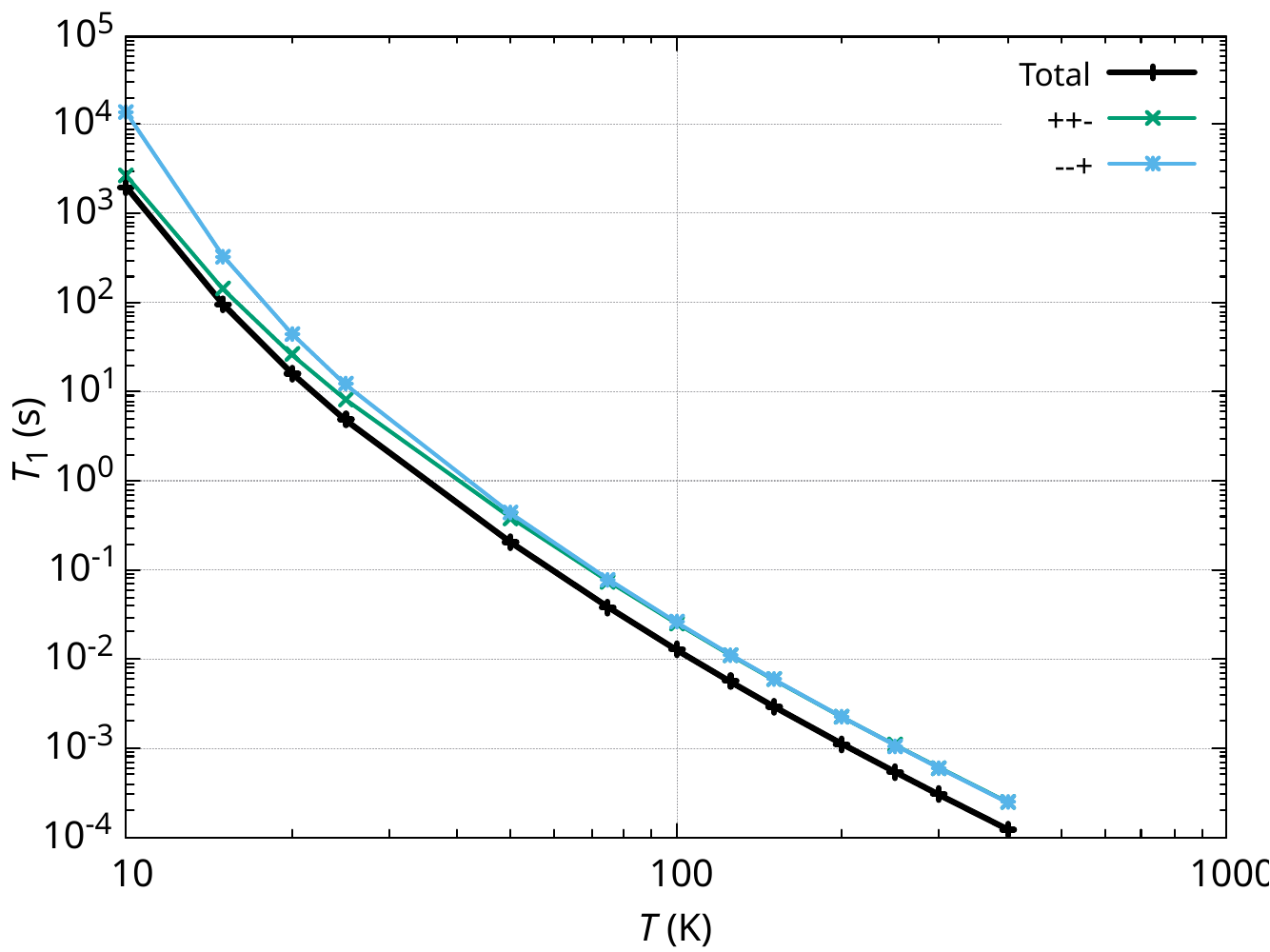}
    \caption{The two major contributions in the three-phonon process, the double emission / single absorption denoted by $R^{++-}$ (green line) and double absorption / single emission denoted by $R^{--+}$ (cyan line) are reported along with the sum of all eight contributions to $R^{(6)}$ (thick black line) as given in Eq. \ref{R6-total}.}
    \label{fig-T1-allW}
\end{figure}

We then proceed to disentangle which, if any in particular, specific three-phonon process determines $R^{(6)}$. Fig. \ref{fig-T1-allW} reports the $T_1$ predicted by each one of the eight contributions to $R^{(6)}$ and shows that the double emission / single absorption ($++-$) and double absorption / single emission ($--+$), with $\hbar\omega_\alpha < \hbar\omega_\beta < \hbar\omega_\gamma$, are the two most, and equally, important contributions to relaxation. In the first process, one high-energy phonon is absorbed, while two lower-energy phonons are emitted. In the second case, the opposite happens, with two low-energy phonons absorbed and one high-energy phonon emitted. This finding is consistent with simple considerations about energy conservation. Since the Zeeman energy is negligible with respect to the vibrational one, the total energy exchanged among the three phonons will need to approximately add up to zero. This immediately discards triple absorption/emission as a possible mechanism. On the other hand, since Eq. \ref{R6-total} is taken over $\hbar\omega_\alpha < \hbar\omega_\beta < \hbar\omega_\gamma$, a condition $ \pm \hbar\omega_\alpha \pm  \hbar\omega_\beta \mp \hbar\omega_\gamma =0 $ is the most likely to lead to energy conservation.

\begin{figure}
    \centering    
\includegraphics[width=\linewidth]{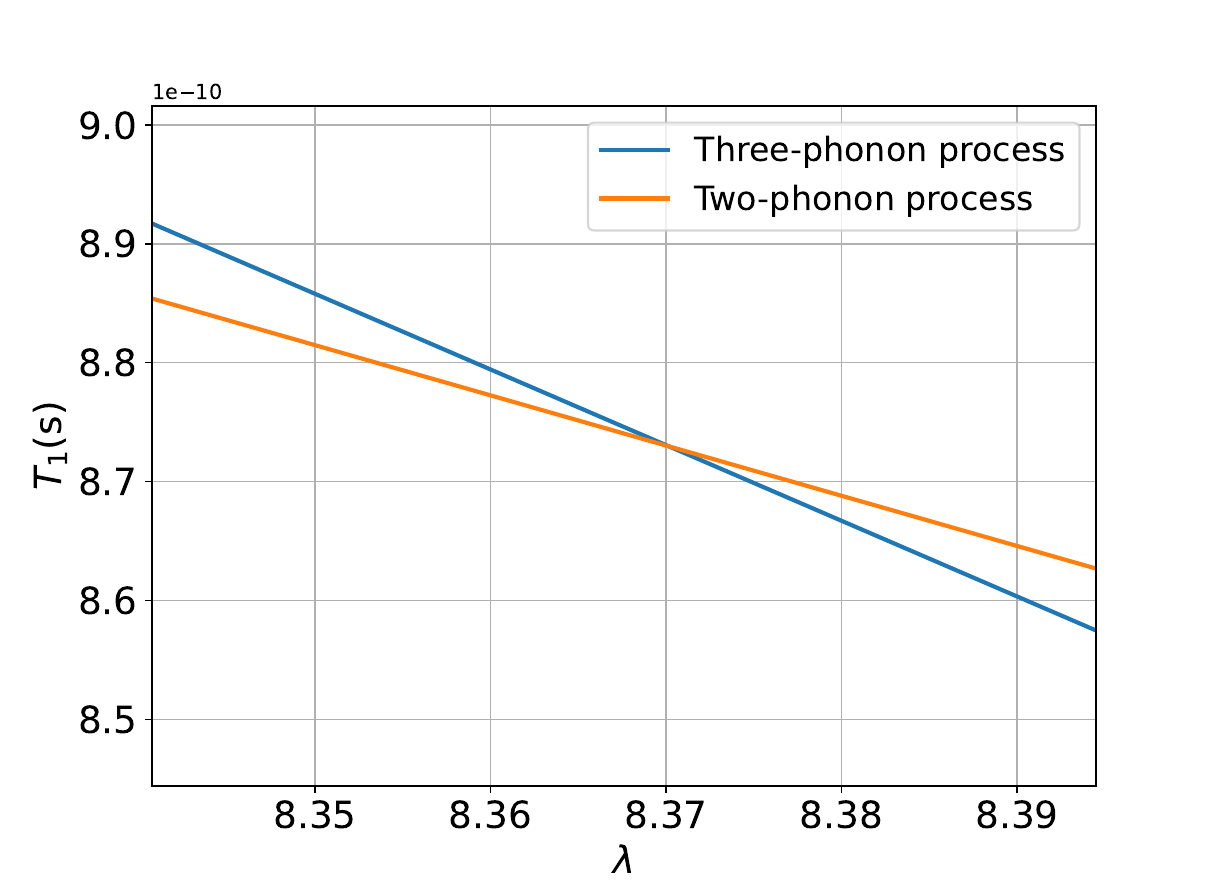}
    \caption{$T_1$ is reported as a function of the scaling factor $\lambda$. At 300 K, the crossover between two-phonon (orange line) and three-phonon (blue line) processes occurs when $\lambda \sim 8.37$. }
    \label{fig-T1vsc}
\end{figure}

Lastly, we explore the conditions under which three-phonon processes might become the leading relaxation mechanism at room temperature. To do so, we homogeneously rescale the vibronic coupling of CrN(pyrdtc)$_{\mathrm 2}$ by a factor $\lambda$ and compute $T_1$ at 300 K as a function of this parameter. Results, reported in Fig.  \ref{fig-T1vsc}, show that a crossover in efficiency between $R^{(4)}$ and $R^{(6)}$ occurs for $\lambda \sim 8.37$ in virtue of the fact that the former scales as $\lambda^4$ and the latter as $\lambda^6$.

\section*{Discussion and Conclusions}

The simulation of the temperature dependence of spin relaxation for both two-phonon and three-phonon processes confirms that the experimental data exhibit good agreement with the two-phonon theory \cite{mariano2025role}, whereas the three-phonon processes are found to be significantly less efficient until temperatures largely exceed 300 K. This result represents an unprecedented confirmation of the validity of the weak-coupling approximation in spin-lattice relaxation, which implies that only the lowest-order terms in a perturbative expansion are necessary to model the relaxation dynamics of this class of spin systems. Interestingly, results also show that a crossover between two- and three-phonon mediated relaxation at 300 K would occur by increasing spin-phonon coupling by a relatively small factor of $\sim 8$. Considering the fact that molecules like the one discussed here are among those with the longest $T_1$ at room temperature, such a crossover might be observable in complexes with shorter $T_1$, where stronger system-environment couplings can be expected. In addition to verifying the validity of the weak coupling assumption and establishing a quantitative way to estimate the onset of strong-coupling effects, we anticipate that the present results will complement non-perturbative methods \cite{anto2021strong, mcconnell2022strong} in modelling spin dynamics in the strong spin-phonon coupling regimes. The latter are known to arise in spin systems when interacting with cavities and nanostructures. For instance, single-molecule magnets (SMMs) coupled to carbon nanotubes, a potential building block for nanoelectromechanical systems, have shown evidence of strong spin-phonon coupling \cite{ganzhorn2013strong}.
Other prominent examples of strong coupling include 
spin-nanomechanical hybrid devices with NV centers \cite{li2016hybrid}, electron spin-embedded optomechanical crystal cavities \cite{peng2025hybrid} and two-dimensional magnetic materials \cite{tian2016magneto, pawbake2023raman}.

In conclusion, we have expanded spin-lattice relaxation theory to the sixth order and accounted for three-phonon processes. The ab initio implementation of this formalism confirms the validity of the weak-coupling approximation for spin-phonon interactions in vdW crystals of isotropic spin-1/2 molecules. On the other hand, the formalism is general and valid for any quantum system interacting with a Markovian phononic bath, paving the way to the exploration of strong-coupling effects in more complex spin, electronic, and quantum systems.

\vspace{0.2cm}
\noindent
\textbf{Acknowledgements and Funding}\\
This project has received funding from the European Research Council (ERC) under the European Union’s Horizon 2020 research and innovation programme (grant agreement No. [948493]). Computational resources were provided by the Trinity College Research IT and the Irish Centre for High-End Computing (ICHEC).

\vspace{0.2cm}
\noindent
\textbf{Data Availability}\\
All the spin dynamics simulations discussed in the present manuscript have been carried out with the software MolForge, which can be obtained at \textit{github.com/LunghiGroup/MolForge}. All data needed to evaluate the conclusions in the paper are present in the paper and/or the Supplementary Materials. Additional data related to this paper may be requested from the authors.

\vspace{0.2cm}
\noindent
\textbf{Conflict of interests}\\
The authors declare that they have no competing interests.

\bibliographystyle{naturemag}
\bibliography{reference}

@article{affronte2009molecular,
  title={Molecular nanomagnets for information technologies},
  author={Affronte, Marco},
  journal={Journal of Materials Chemistry},
  volume={19},
  number={12},
  pages={1731--1737},
  year={2009},
  publisher={Royal Society of Chemistry}
}

@article{moseley2018spin,
  title={Spin--phonon couplings in transition metal complexes with slow magnetic relaxation},
  author={Moseley, Duncan H and Stavretis, Shelby E and Thirunavukkuarasu, Komalavalli and Ozerov, Mykhaylo and Cheng, Yongqiang and Daemen, Luke L and Ludwig, Jonathan and Lu, Zhengguang and Smirnov, Dmitry and Brown, Craig M and others},
  journal={Nature Communications},
  volume={9},
  number={1},
  pages={2572},
  year={2018},
  publisher={Nature Publishing Group UK London}
}

@article{orbach1961spin,
  title={Spin-lattice relaxation in rare-earth salts},
  author={Orbach, R},
  journal={Proceedings of the Royal Society of London. Series A. Mathematical and Physical Sciences},
  volume={264},
  number={1319},
  pages={458--484},
  year={1961},
  publisher={The Royal Society London}
}

@article{gugler2018ab,
  title={Ab initio calculation of the spin lattice relaxation time T 1 for nitrogen-vacancy centers in diamond},
  author={Gugler, Johannes and Astner, Thomas and Angerer, Andreas and Schmiedmayer, J{\"o}rg and Majer, Johannes and Mohn, Peter},
  journal={Physical Review B},
  volume={98},
  number={21},
  pages={214442},
  year={2018},
  publisher={APS}
}

@article{mondal2023spin,
  title={Spin-phonon decoherence in solid-state paramagnetic defects from first principles},
  author={Mondal, Sourav and Lunghi, Alessandro},
  journal={npj Computational Materials},
  volume={9},
  number={1},
  pages={120},
  year={2023},
  publisher={Nature Publishing Group UK London}
}

@article{khaetskii2000spin,
  title={Spin relaxation in semiconductor quantum dots},
  author={Khaetskii, Alexander V and Nazarov, Yuli V},
  journal={Physical Review B},
  volume={61},
  number={19},
  pages={12639},
  year={2000},
  publisher={APS}
}

@article{gong2019two,
  title={Two-dimensional magnetic crystals and emergent heterostructure devices},
  author={Gong, Cheng and Zhang, Xiang},
  journal={Science},
  volume={363},
  number={6428},
  pages={eaav4450},
  year={2019},
  publisher={American Association for the Advancement of Science}
}

@article{bloch1946nuclear,
  title={Nuclear induction},
  author={Bloch, Felix},
  journal={Physical review},
  volume={70},
  number={7-8},
  pages={460},
  year={1946},
  publisher={APS}
}

@article{birk2003atom,
  title={Atom transfer as a preparative tool in coordination chemistry. Synthesis and characterization of Cr (V) nitrido complexes of bidentate ligands},
  author={Birk, Torben and Bendix, Jesper},
  journal={Inorganic chemistry},
  volume={42},
  number={23},
  pages={7608--7615},
  year={2003},
  publisher={ACS Publications}
}

@article{nakajima1958quantum,
  title={On quantum theory of transport phenomena: Steady diffusion},
  author={Nakajima, Sadao},
  journal={Progress of Theoretical Physics},
  volume={20},
  number={6},
  pages={948--959},
  year={1958},
  publisher={Oxford University Press}
}

@article{zwanzig1960ensemble,
  title={Ensemble method in the theory of irreversibility},
  author={Zwanzig, Robert},
  journal={The Journal of Chemical Physics},
  volume={33},
  number={5},
  pages={1338--1341},
  year={1960},
  publisher={American Institute of Physics}
}

@article{chaturvedi1979time,
  title={Time-convolutionless projection operator formalism for elimination of fast variables. Applications to Brownian motion},
  author={Chaturvedi, S and Shibata, F},
  journal={Zeitschrift f{\"u}r Physik B Condensed Matter},
  volume={35},
  number={3},
  pages={297--308},
  year={1979},
  publisher={Springer}
}

@article{shibata1977generalized,
  title={A generalized stochastic liouville equation. Non-Markovian versus memoryless master equations},
  author={Shibata, Fumiaki and Takahashi, Yoshinori and Hashitsume, Natsuki},
  journal={Journal of Statistical Physics},
  volume={17},
  number={4},
  pages={171--187},
  year={1977},
  publisher={Springer}
}

@article{shibata1980expansion,
  title={Expansion formulas in nonequilibrium statistical mechanics},
  author={Shibata, Fumiaki and Arimitsu, Toshihico},
  journal={Journal of the Physical Society of Japan},
  volume={49},
  number={3},
  pages={891--897},
  year={1980},
  publisher={The Physical Society of Japan}
}

@article{breuer1999stochastic,
  title={Stochastic wave-function method for non-Markovian quantum master equations},
  author={Breuer, Heinz-Peter and Kappler, Bernd and Petruccione, Francesco},
  journal={Physical Review A},
  volume={59},
  number={2},
  pages={1633},
  year={1999},
  publisher={APS}
}

@article{leggett1987dynamics,
  title={Dynamics of the dissipative two-state system},
  author={Leggett, Anthony J and Chakravarty, SDAFMGA and Dorsey, Alan T and Fisher, Matthew PA and Garg, Anupam and Zwerger, Wilhelm},
  journal={Reviews of Modern Physics},
  volume={59},
  number={1},
  pages={1},
  year={1987},
  publisher={APS}
}

@book{bruus2004many,
  title={Many-body quantum theory in condensed matter physics: an introduction},
  author={Bruus, Henrik and Flensberg, Karsten},
  year={2004},
  publisher={Oxford university press}
}

@article{waller1932magnet,
  title={{\"U}ber die Magnetisierung von paramagnetischen Kristallen in Wechselfeldern},
  author={Waller, I},
  journal={Zeitschrift f{\"u}r Physik},
  volume={79},
  number={5},
  pages={370--388},
  year={1932},
  publisher={Springer}
}

@incollection{gorter1957chapter,
  title={Chapter IX Paramagnetic Relaxation},
  author={Gorter, CJ},
  booktitle={Progress in Low Temperature Physics},
  volume={2},
  pages={266--291},
  year={1957},
  publisher={Elsevier}
}

@article{heitler1936time,
  title={Time effects in the magnetic cooling method},
  author={Heitler, W and Teller, Edward},
  journal={Proceedings of the Royal Society of London. Series A-Mathematical and Physical Sciences},
  volume={155},
  number={886},
  pages={629--639},
  year={1936},
  publisher={The Royal Society London}
}

@article{fierz1938theory,
  title={The theory of the susceptibility of paramagnetic alaun in transitional fields},
  author={Fierz, M},
  journal={Physica},
  volume={5},
  pages={433},
  year={1938}
}

@article{kronig1939mechanism,
  title={On the mechanism of paramagnetic relaxation},
  author={Kronig, R de L},
  journal={Physica},
  volume={6},
  number={1},
  pages={33--43},
  year={1939},
  publisher={Elsevier}
}

@article{van1939jahn,
  title={The Jahn-Teller effect and crystalline Stark splitting for clusters of the form XY6},
  author={Van Vleck, JH},
  journal={The Journal of Chemical Physics},
  volume={7},
  number={1},
  pages={72--84},
  year={1939},
  publisher={American Institute of Physics}
}

@article{van1940paramagnetic,
  title={Paramagnetic relaxation times for titanium and chrome alum},
  author={Van Vleck, JH},
  journal={Physical Review},
  volume={57},
  number={5},
  pages={426},
  year={1940},
  publisher={APS}
}

@article{redfield1957theory,
  title={On the theory of relaxation processes},
  author={Redfield, Alfred G},
  journal={IBM Journal of Research and Development},
  volume={1},
  number={1},
  pages={19--31},
  year={1957},
  publisher={IBM}
}

@article{lindblad1976generators,
  title={On the generators of quantum dynamical semigroups},
  author={Lindblad, Goran},
  journal={Communications in mathematical physics},
  volume={48},
  number={2},
  pages={119--130},
  year={1976},
  publisher={Springer}
}

@article{gorini1976completely,
  title={Completely positive dynamical semigroups of N-level systems},
  author={Gorini, Vittorio and Kossakowski, Andrzej and Sudarshan, Ennackal Chandy George},
  journal={Journal of Mathematical Physics},
  volume={17},
  number={5},
  pages={821--825},
  year={1976},
  publisher={American Institute of Physics}
}

@book{alicki2007quantum,
  title={Quantum dynamical semigroups and applications},
  author={Alicki, Robert and Lendi, Karl},
  year={2007},
  publisher={Springer}
}

@article{kraus2008preparation,
  title={Preparation of entangled states by quantum Markov processes},
  author={Kraus, Barbara and B{\"u}chler, Hans P and Diehl, Sebastian and Kantian, Adrian and Micheli, Andrea and Zoller, Peter},
  journal={Physical Review A—Atomic, Molecular, and Optical Physics},
  volume={78},
  number={4},
  pages={042307},
  year={2008},
  publisher={APS}
}

@article{sieberer2016keldysh,
  title={Keldysh field theory for driven open quantum systems},
  author={Sieberer, Lukas M and Buchhold, Michael and Diehl, Sebastian},
  journal={Reports on Progress in Physics},
  volume={79},
  number={9},
  pages={096001},
  year={2016},
  publisher={IOP Publishing}
}

@article{wang2011quantum,
  title={Quantum algorithm for simulating the dynamics of an open quantum system},
  author={Wang, Hefeng and Ashhab, Sahel and Nori, Franco},
  journal={Physical Review A—Atomic, Molecular, and Optical Physics},
  volume={83},
  number={6},
  pages={062317},
  year={2011},
  publisher={APS}
}

@article{anto2021strong,
  title={Strong coupling effects in quantum thermal transport with the reaction coordinate method},
  author={Anto-Sztrikacs, Nicholas and Segal, Dvira},
  journal={New Journal of Physics},
  volume={23},
  number={6},
  pages={063036},
  year={2021},
  publisher={IOP Publishing}
}

@article{mcconnell2022strong,
  title={Strong coupling in thermoelectric nanojunctions: A reaction coordinate framework},
  author={McConnell, Conor and Nazir, Ahsan},
  journal={New Journal of Physics},
  volume={24},
  number={2},
  pages={025002},
  year={2022},
  publisher={IOP Publishing}
}

@article{ganzhorn2013strong,
  title={Strong spin--phonon coupling between a single-molecule magnet and a carbon nanotube nanoelectromechanical system},
  author={Ganzhorn, Marc and Klyatskaya, Svetlana and Ruben, Mario and Wernsdorfer, Wolfgang},
  journal={Nature nanotechnology},
  volume={8},
  number={3},
  pages={165--169},
  year={2013},
  publisher={Nature Publishing Group UK London}
}

@article{li2016hybrid,
  title={Hybrid quantum device with nitrogen-vacancy centers in diamond coupled to carbon nanotubes},
  author={Li, Peng-Bo and Xiang, Ze-Liang and Rabl, Peter and Nori, Franco},
  journal={Physical review letters},
  volume={117},
  number={1},
  pages={015502},
  year={2016},
  publisher={APS}
}

@article{peng2025hybrid,
  title={Hybrid spin-phonon architecture for scalable solid-state quantum nodes},
  author={Peng, Ruoming and Wu, Xuntao and Wang, Yang and Zhang, Jixing and Geng, Jianpei and Dasari, Durga Bhaktavatsala Rao and Cleland, Andrew N and Wrachtrup, J{\"o}rg},
  journal={npj Quantum Information},
  volume={11},
  number={1},
  pages={176},
  year={2025},
  publisher={Nature Publishing Group UK London}
}

@article{tian2016magneto,
  title={Magneto-elastic coupling in a potential ferromagnetic 2D atomic crystal},
  author={Tian, Yao and Gray, Mason J and Ji, Huiwen and Cava, Robert J and Burch, Kenneth S},
  journal={2D Materials},
  volume={3},
  number={2},
  pages={025035},
  year={2016},
  publisher={IOP Publishing}
}

@article{pawbake2023raman,
  title={Raman scattering signatures of strong spin-phonon coupling in the bulk magnetic van der Waals material CrSBr},
  author={Pawbake, Amit and Pelini, Thomas and Wilson, Nathan P and Mosina, Kseniia and Sofer, Zdenek and Heid, Rolf and Faugeras, Clement},
  journal={Physical Review B},
  volume={107},
  number={7},
  pages={075421},
  year={2023},
  publisher={APS}
}

@article{lunghi2019phonons,
  title={How do phonons relax molecular spins?},
  author={Lunghi, Alessandro and Sanvito, Stefano},
  journal={Science advances},
  volume={5},
  number={9},
  pages={eaax7163},
  year={2019},
  publisher={American Association for the Advancement of Science}
}

@article{lunghi2022toward,
  title={Toward exact predictions of spin-phonon relaxation times: An ab initio implementation of open quantum systems theory},
  author={Lunghi, Alessandro},
  journal={Science Advances},
  volume={8},
  number={31},
  pages={eabn7880},
  year={2022},
  publisher={American Association for the Advancement of Science}
}

@article{lunghi2020limit,
  title={The limit of spin lifetime in solid-state electronic spins},
  author={Lunghi, Alessandro and Sanvito, Stefano},
  journal={The Journal of Physical Chemistry Letters},
  volume={11},
  number={15},
  pages={6273--6278},
  year={2020},
  publisher={ACS Publications}
}

@article{breuer2016colloquium,
  title={Colloquium: Non-Markovian dynamics in open quantum systems},
  author={Breuer, Heinz-Peter and Laine, Elsi-Mari and Piilo, Jyrki and Vacchini, Bassano},
  journal={Reviews of Modern Physics},
  volume={88},
  number={2},
  pages={021002},
  year={2016},
  publisher={APS}
}

@article{breuer2001time,
  title={The time-convolutionless projection operator technique in the quantum theory of dissipation and decoherence},
  author={Breuer, Heinz-Peter and Kappler, Bernd and Petruccione, Francesco},
  journal={Annals of Physics},
  volume={291},
  number={1},
  pages={36--70},
  year={2001},
  publisher={Elsevier}
}

@book{schieve2009quantum,
  title={Quantum statistical mechanics},
  author={Schieve, William C and Horwitz, Lawrence P},
  year={2009},
  publisher={Cambridge University Press}
}

@article{akera1999coulomb,
  title={Coulomb staircase and total spin in quantum dots},
  author={Akera, Hiroshi},
  journal={Physical Review B},
  volume={60},
  number={15},
  pages={10683},
  year={1999},
  publisher={APS}
}

@article{golovach2004transport,
  title={Transport through a double quantum dot in the sequential tunneling and cotunneling regimes},
  author={Golovach, Vitaly N and Loss, Daniel},
  journal={Physical Review B—Condensed Matter and Materials Physics},
  volume={69},
  number={24},
  pages={245327},
  year={2004},
  publisher={APS}
}

@article{gross1999covariant,
  title={Covariant representations of the relativistic Brueckner T-matrix and the nuclear matter problem},
  author={Gross-Boelting, T and Fuchs, Ch and Faessler, Amand},
  journal={Nuclear Physics A},
  volume={648},
  number={1-2},
  pages={105--137},
  year={1999},
  publisher={Elsevier}
}

@article{soma2008medium,
  title={In-medium T matrix for nuclear matter with three-body forces: Binding energy and single-particle properties},
  author={Soma, V and Bo{\.z}ek, P},
  journal={Physical Review C—Nuclear Physics},
  volume={78},
  number={5},
  pages={054003},
  year={2008},
  publisher={APS}
}

@article{koch2004thermopower,
  title={Thermopower of single-molecule devices},
  author={Koch, Jens and Von Oppen, Felix and Oreg, Yuval and Sela, Eran},
  journal={Physical Review B—Condensed Matter and Materials Physics},
  volume={70},
  number={19},
  pages={195107},
  year={2004},
  publisher={APS}
}

@article{jorn2006theory,
  title={Theory of current-induced dynamics in molecular-scale devices},
  author={Jorn, Ryan and Seideman, Tamar},
  journal={The Journal of chemical physics},
  volume={124},
  number={8},
  year={2006},
  publisher={AIP Publishing}
}

@article{timm2011time,
  author={Timm, Carsten},
  journal={Physical Review B—Condensed Matter and Materials Physics},
  title={Time-convolutionless master equation for quantum dots: Perturbative expansion to arbitrary order},
  volume={83},
  pages={115416},
  year={2011},
  publisher={APS}
}

@article{mariano2025role,
  author={Mariano et al., Lorenzo A.},
  title={The role of electronic excited states in the spin-lattice relaxation of spin-1/2 molecules},
  journal={Science Advances},
  volume={11},
  pages={eadr0168},
  year={2025},
  publisher={American Association for the Advancement of Science}
}

@article{mariano2024vibr,
author = {Mariano, Lorenzo A. and Mondal, Sourav and Lunghi, Alessandro},
title = {Spin-Vibronic Dynamics in Open-Shell Systems beyond the Spin Hamiltonian Formalism},
journal = {Journal of Chemical Theory and Computation},
volume = {20},
number = {1},
pages = {323-332},
year = {2024},
publisher={ACS Publications}
}

@article{eaton2025anisotropy,
author = {Eaton, Sandra S. and Yamabayashi, Tsutomu and Horii, Yoji and Yamashita, Masahiro and Eaton, Gareth R.},
title = {Anisotropy of Spin–Lattice Relaxation Time (T1) for Oxo-Vanadium(IV) and Nitrido Chromium(V) Porphyrins},
journal = {Journal of the American Chemical Society},
volume = {147},
number = {16},
pages = {13815-13823},
year = {2025},
publisher={ACS Publications}
}

@article{kazmi2025esr,
author = {Kazmierczak, Nathanael P. and Xia, Kay T. and Sutcliffe, Erica and Aalto, Jonathan P. and Hadt, Ryan G.},
title = {A Spectrochemical Series for Electron Spin Relaxation},
journal = {Journal of the American Chemical Society},
volume = {147},
number = {3},
pages = {2849-2859},
year = {2025},
publisher={ACS Publications}
}

@article{lunghi2025fourth,
  title={Fourth-order quantum master equations reveal that spin-phonon decoherence undercuts long magnetization relaxation times in single-molecule magnets},
  author={Lunghi, Alessandro},
  journal={arXiv preprint arXiv:2507.20716},
  year={2025}
}

\end{document}